\documentclass[9pt,twocolumn,twoside]{pnas-new}

\usepackage{hyperref}

\templatetype{pnasresearcharticle} 

\title{Circumpolar ocean stability on Mars 3 Gy ago}

\author[a,b,1,2]{Fr{\'e}d{\'e}ric Schmidt}
\author[c,d,e,1]{Michael J. Way} 
\author[a]{Fran{\c c}ois Costard}
\author[a,b,f]{Sylvain Bouley}
\author[a]{Antoine S{\'e}journ{\'e}}
\author[c,g]{Igor Aleinov}

\affil[a]{Universit{\'e} Paris-Saclay, CNRS, GEOPS, 91405, ORSAY, FRANCE}
\affil[b]{Institut Universitaire de France (IUF)}
\affil[c]{NASA/Goddard Institute for Space Studies, 2880 Broadway, NY, NY 10025, USA}
\affil[d]{GSFC Sellers Exoplanet Environments Collaboration, Greenbelt, MD, USA}
\affil[e]{Theoretical Astrophysics, Department of Physics and Astronomy, Uppsala University, Uppsala, SWEDEN}
\affil[f]{IMCCE -- Observatoire de Paris, CNRS-UMR 8028, Paris, France}
\affil[g]{Center for Climate Systems Research, Columbia University, New York, NY 10025, USA}

\leadauthor{Schmidt} 

\significancestatement{The current Martian climate is not habitable and far from Earth's climate. At the same time that Life spread on Earth (3 Gy ago), the Red Planet was possibly more similar to our Blue Planet. Our model includes a coupled model with dynamic ocean and atmosphere including a hydrological cycle and a simplified glacier mass flux scheme. We show that an ocean is stable in agreement with interpretations of the surface geological records.}

\authorcontributions{Please provide details of author contributions here.}
\authordeclaration{Please declare any competing interests here.}
\equalauthors{\textsuperscript{1}F.S. contributed equally to this work with M.J.W.}
\correspondingauthor{\textsuperscript{2}To whom correspondence should be addressed. E-mail: frederic.schmidt@@universite-paris-saclay.fr}


\begin{abstract}
What was the nature of the Late Hesperian climate? Warm and wet or cold and dry? Formulated this way the question leads to an apparent paradox since both options seem implausible. A warm and wet climate would have produced extensive fluvial erosion but few valley networks have been observed at the age of the late Hesperian. A too cold climate would have kept any northern ocean frozen most of the time. A moderate cold climate would have transferred the water from the ocean to the land in the form of snow and ice. But this would prevent tsunami formation, for which there is some evidence. Here, we provide new insights from numerical climate simulations in agreement with surface geological features to demonstrate that the Martian climate could have been both cold and wet. Using an advanced General Circulation Model (GCM), we demonstrate that an ocean can be stable, even if the Martian mean surface temperature is lower than 0$^\circ$C. Rainfall is moderate near the shorelines and in the ocean. The southern plateau is mostly covered by ice with a mean temperature below 0$^\circ$C and a glacier return flow back to the ocean. This climate is achieved with a 1 bar CO$_2$ dominated atmosphere with 10\% H$_2$. Under this scenario 3 Ga, the geologic evidence of a shoreline and tsunami deposits along the ocean/land dichotomy are compatible with ice sheets and glacial valleys in the southern highlands.
\end{abstract}

\dates{This manuscript was compiled on \today}
\doi{\url{www.pnas.org/cgi/doi/10.1073/pnas.XXXXXXXXXX}}

\begin{document}

\maketitle
\thispagestyle{firststyle}
\ifthenelse{\boolean{shortarticle}}{\ifthenelse{\boolean{singlecolumn}}{\abscontentformatted}{\abscontent}}{}

\dropcap{T}he possibility of a late Martian ocean is a topic of debate with strong implications on the habitability of the Red Planet \cite{Baker_Ancientoceansice_N1991}.
A recent review \cite{Dickeson_Martianoceans_AG2020} discusses this controversy.
There is evidence of Martian paleo-shorelines \cite{Parker_CoastalgeomorphologyMartian_JoGR1993} in Deuteronilus Mensae (sometimes noted contact No 2) in a geometry closely corresponding to the current equipotential height \cite{Head_PossibleAncientOceans_Science1999}. The Deuteronilus shoreline seems to have formed during the last stage of the true polar wander induced by Tharsis \cite{Citron_Timingofoceans_N2018}. A northern ocean is also supported by specific radar properties \cite{Mouginot_Dielectricmapof_GRL2012}, smooth surface roughness \cite{Kreslavsky_SlopeMOLA_JGR2000} and by a fractal analysis of the topography  \cite{Saberi_EvidenceAncientSea_TAJ2020}. Crater count dating of the Vastitas Borealis Formation near Deuteronilus Mensae is 3.5 Ga \cite{Ivanov_TopographyDeuteroniluscontact_PaSS2017} but the ocean may have been more recent. Detailed studies in Kasei Valles imply that such an ocean rose in elevation ($\sim$ 1000 m) between ca. 3.6 Ga and 3.2 Ga \cite{Duran_KaseiVallesMars_SR2020}. Along this shoreline, candidate tsunami deposits have been identified at an age of 3 Ga \cite{Rodriguez_Tsunamiwavesextensively_SR2016,Costard_Modelingtsunamipropagation_JoGRP2017} with at least two impact events. The Lomonosov crater morphology is coherent with an impact in shallow water, at the very same age of the tsunami deposits \cite{Costard_LomonosovCraterImpact_JoGRP2019} and it is thus the most probable source.

The stability of an ocean in a warm and wet scenario, even transient, at 3 Ga has been questioned since intense and widespread rainfall in such a scenario appeared inconsistent with the few observed dendritic valley networks from this time  \cite{Wordsworth_ClimateEarlyMars_ARoEaPS2016, Turbet_paradoxesLateHesperian_SR2019}. In previous work, a cold and wet Mars seemed impossible since the long term stability of an ocean (open or ice covered) in such a scenario has never been achieved by a three dimensional General Circulation Model (3D-GCM) \cite{Forget_3Dmodellingearly_I2013, Wordsworth_Globalmodellingearly_I2013, 
Wordsworth_ClimateEarlyMars_ARoEaPS2016, Turbet_3Dmodellingclimatic_I2017, Turbet_paradoxesLateHesperian_SR2019, Kite_2021PNAS}. Mainly the water was found to accumulate in the form of ice in the southern highlands, but no glacier/ice sheet processes were considered in these studies.

Cool and wet scenarios in the case of a faint young sun (at the Noachian-Hesperian boundary at 3.6 Ga) demonstrate that the ocean freezes for pressures below 1 bar and yet predict intense rainfall when the global average temperature is higher than 0$^\circ$C \cite{Kamada_coupledatmospherehydrosphereglobal_I2020}. Nevertheless, liquid brines are possible in this case but only with significant anti-freezing properties \cite{Fairen_ColdWetMars_Icarus2010}. An investigation with 2D simulations for early Mars argues for a warm and semi-arid early Mars \cite{Ramirez_ClimateSimulationsEarly_JoGRP2020} but stresses the need for a coupled ocean/atmosphere model.

\section*{3D climate model}

We present fully coupled ocean/atmosphere 3-D GCM simulations based on ROCKE-3D \citep[][hereafter R3D]{Way2017}, which is based upon a parent Earth Climate Model known as ModelE2 \cite{Schmidt2014}. R3D allows us to estimate the interaction between atmosphere/ocean circulation but also encompasses a surface hydrological scheme. We assume the solar luminosity to be $\sim$ 79\% \cite{Gough1981} of its current value. Hence at 3 Ga the flux at Mars was set to 452.8 W.m$^{-2}$ in the GCM. The total water budget is 150 m GEL (Global Equivalent Layer) to fill the ocean basin in the north and Hellas up to -3900 m. Our model includes a dynamic fully-coupled ocean while sea ice is also interactive and determined by the salinity and corresponding freezing point of water \cite{Way2017}. The salinity was set to modern Earth ocean values at model start given the lack of constraints for ancient Mars, but R3D is capable of exploring a range of salinities \cite{DelGenio2019}. In addition, we include a glacier flux in R3D to simulate the return flow to the ocean. This scheme allows us to estimate the glacier mass balance, but glacier thickness (including total mass) and flowing path cannot be constrained by our model. In any case, the water budget must be compatible with D/H measurements that imply 100-300 m GEL in the Hesperian \cite{Mahaffy_imprintatmosphericevolution_S2015, Scheller_Longtermdrying_S2021}, hence our choice of 150 m GEL.

Four sets of numerical simulations were performed for 0$^\circ$, 20$^\circ$, 40$^\circ$, 60$^\circ$ obliquity, with 1 bar atmospheric pressure \cite{Turbet_paradoxesLateHesperian_SR2019, Kite_GeologicConstraintsEarly_SSR2019}. An obliquity of 40$^\circ$ seems to be the most likely \cite{Laskar_INSOLATIONofMARS_Icarus_2004}. We set eccentricity to 0 since it has only a minor effect on mean annual temperature \cite{Palumbo_EarlyMarsClimate_GRL2018,Palumbo_LateNoachianIcy_I2018}. For 20\% H$_2$ in a CO$_2$ dominated atmosphere the greenhouse effect is efficient enough to keep most of the planet's surface temperature above freezing (Sup. Mat.) as expected \cite{Kamada_coupledatmospherehydrosphereglobal_I2020}. The global mean surface temperature is always greater than 10$^\circ$C with a stable ocean (Northern ocean and Hellas basin sea) including intense rainfall, corresponding to the warm and wet scenario \cite{Wordsworth_Comparisonwarmwet_JoGRP2015}. For 10\% H$_2$ the globally averaged surface temperature is below 0$^\circ$C but the Northern ocean surface temperature remains surprisingly around 7$^\circ$C and thus the ocean is prevented from freezing. The source of H$_2$ may be volcanic outgassing or serpentinization, but such large concentrations are not expected to persist for more than a My \cite{Ramirez_WarmingearlyMars_NG2014, Batalha_TestingearlyMars_I2015, Wordsworth2017, Wordsworth2021}. Nevertheless, volcanic activities in Tharsis and Elysium have been identified to cover very large period of time, from 3.8 Ga to 200 Ma \cite{Neukum_RecentVolcanicANDglacial_Nature2004}. Alternatively H$_2$ from impact degassing has also been proposed \cite{Haberle2019}. The lower limit of H$_2$ concentration necessary to stabilize the ocean will be a subject of a future investigation.

\begin{SCfigure*}[\sidecaptionrelwidth][t]
\centering
\includegraphics[width=15cm]{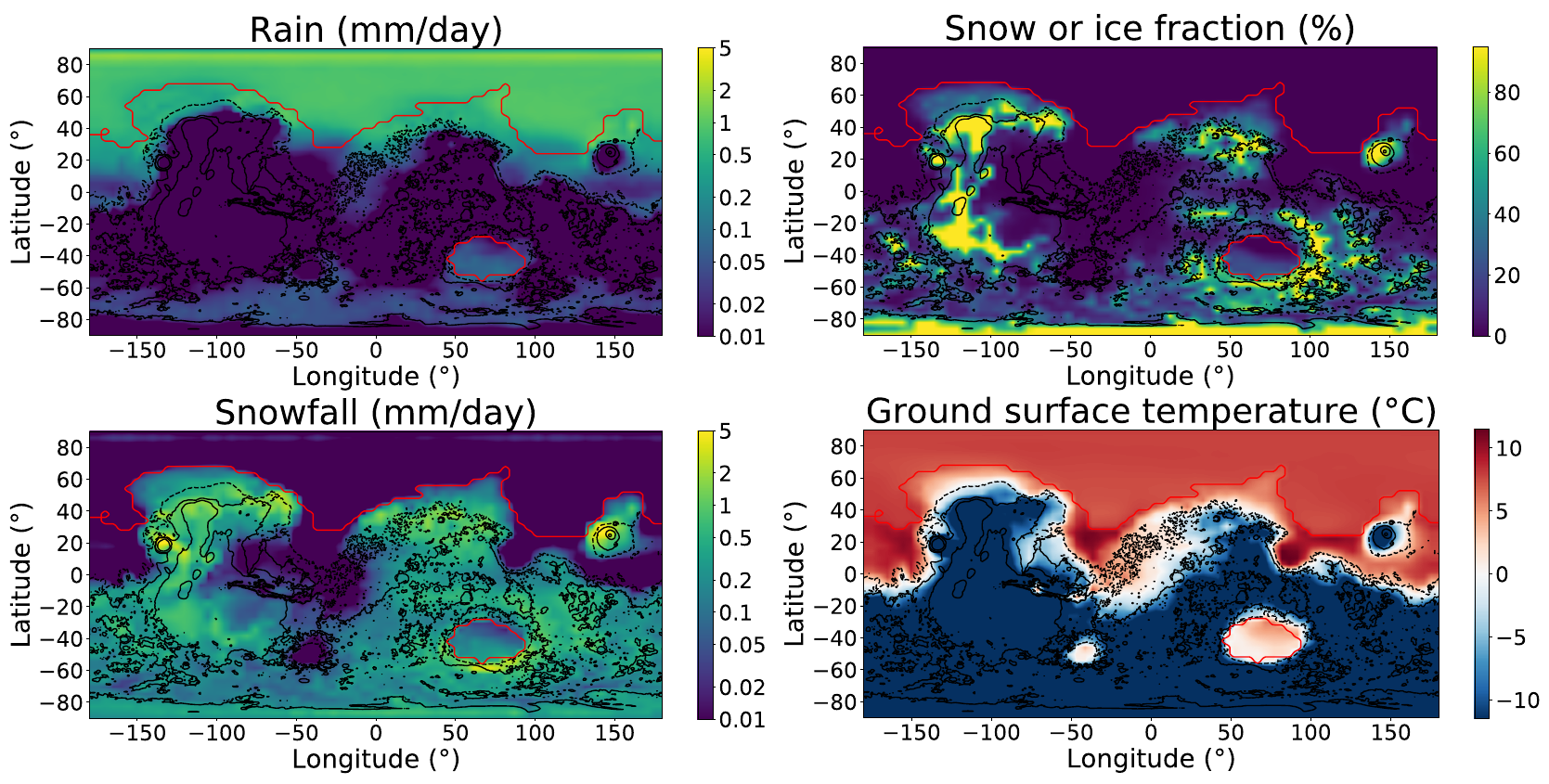}
\caption{3D GCM output at 40$^\circ$ obliquity and H$_2$=10\% for the rain precipitation, snow and ice fraction at the surface, snowfall, sea/ground surface temperature. Black contour lines represent surface elevation level (-2000, 0, 2000 and 8000 meters) and the red contour line is the paleo-shoreline (-3900 meters).}
\label{Fig:GCM-output}
\end{SCfigure*}

Figure \ref{Fig:GCM-output} shows the simulated surface fields averaged over 10 Martian years for H$_2$=10\% and obliquity 40$^\circ$. Interestingly the climate only changes slightly with different obliquities (See Sup. Mat.) likely because the mean ground albedo does not change significantly (e.g. 23\% versus 28\% for 0$^\circ$ versus 60$^\circ$ obliquity). The major effects of water, ice and snow albedo dependence on incidence angles are included (See Materials and Methods). The relatively modest effects of obliquity are probably due to the 1 bar atmosphere that more efficiently transports heat than the present-day thin 6 mbar atmosphere.

\begin{figure}
\center
\includegraphics[viewport=40bp 120bp 930bp 580bp,clip, width=1\columnwidth]{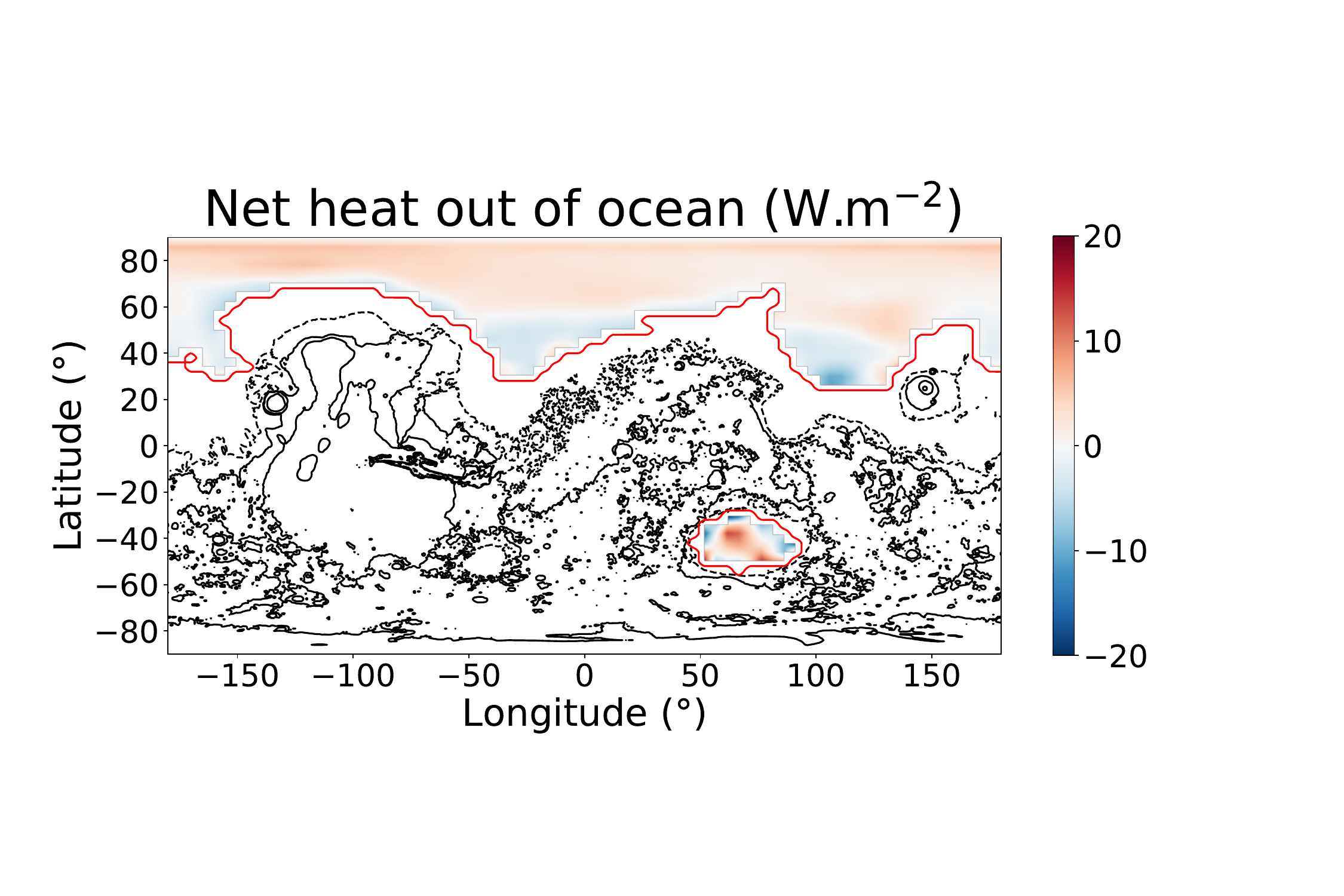}
\includegraphics[width=1\columnwidth]{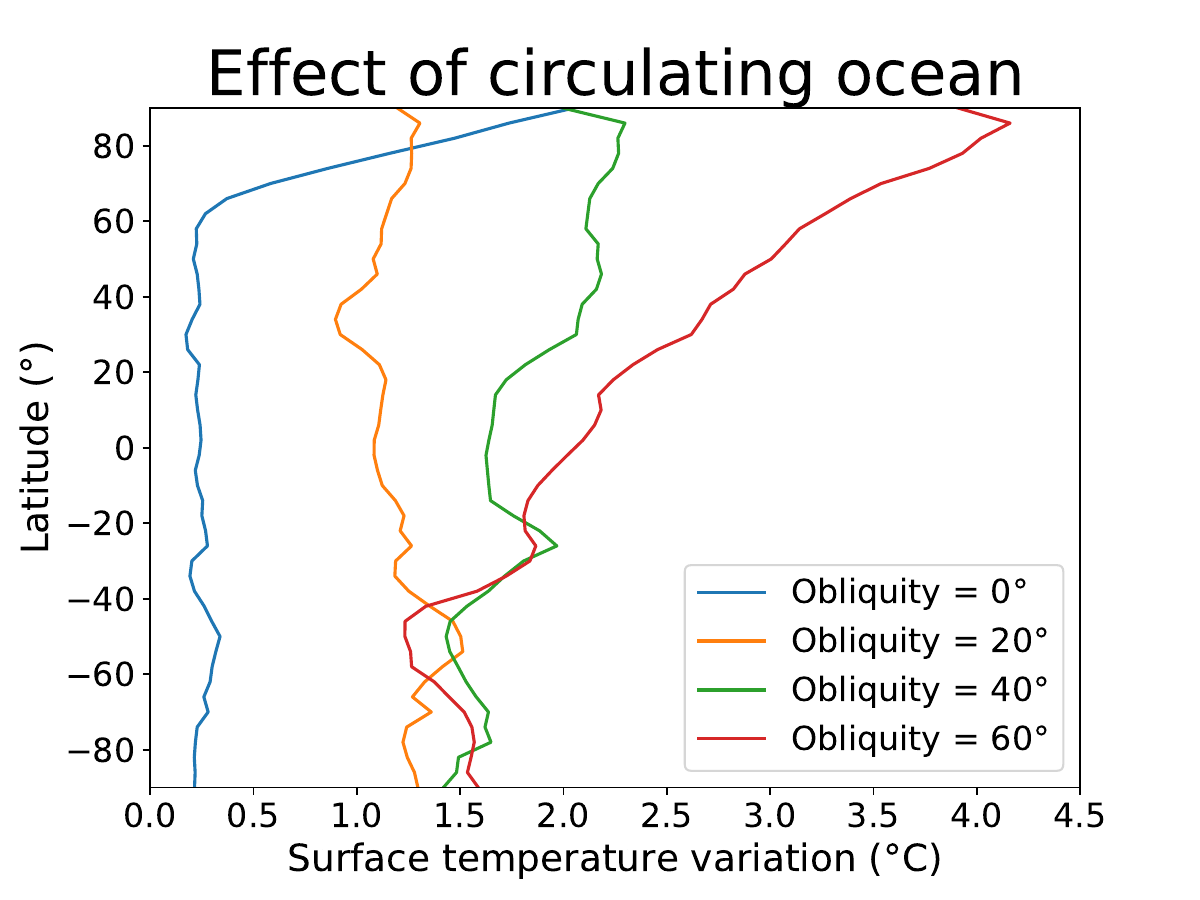}
\caption{(top) Net outward heat flux transported by the ocean for 40$^\circ$ obliquity and H$_2$=10\%. The positive value (up to 15 W.m$^{-2}$) near the North pole indicates that heat goes toward the atmosphere due to ocean circulation. For a slab ocean model, this flux is null. Black contour lines represent surface elevation level and the red contour line is the paleo-shoreline. (bottom) Latitudinal profile of the temperature increase ($T_{circulating}$-$T_{slab}$) due to the circulating ocean, as a function of obliquity. In the slab case, the ocean is assumed stratified but non circulating.}
\label{Fig:GCM-ocean-heat}

\end{figure}

Despite an average planetary surface temperature below 0$^\circ$C, the ocean remains above freezing due to its low altitude and low albedo (Sup. Mat. Table 1). In addition the ability of the ocean to transport heat is responsible for non-negligible higher local temperatures. Figure \ref{Fig:GCM-ocean-heat} top shows the net vertical heat flux contributing to the ocean's stabilization. Ocean gyres are transporting heat poleward, as expected for a fast-rotating planet \cite{Way_ClimatesWarmEarth_TAJSS2018}. The importance of ocean heat transport in a dynamic ocean versus a slab ocean was quantified in \cite{DelGenio2019,Way_ClimatesWarmEarth_TAJSS2018}, and the ability to extract wind driven ocean upwelling related nutrient supply for life from R3D simulations has also recently been demonstrated \cite{Olson2020}. Figure \ref{Fig:GCM-ocean-heat} bottom presents the surface temperature increase due to the active circulation of the ocean as a function of latitude and obliquity. The warming is globally present at all latitudes but locally higher near the northern polar ocean, from 1$^\circ$C up to 4.5$^\circ$C. This effect is larger for extreme obliquities 0$^\circ$ and 60$^\circ$ due to the insolation regime at the poles.

On land, there is a clear boundary at the 0$^\circ$C isotherm which corresponds approximately to the Martian dichotomy. The altitude of the boundary varies from -700 to -3000 meters for obliquity from 0$^\circ$ to 60$^\circ$ (see Sup. Mat.). 

In the high altitude domain, commonly referred to as the "icy highlands", the surface is mostly frozen and snow precipitation is dominant. The extensive accumulation of snow in the highlands can lead to the formation of significant ice sheets that may flow down to the Northern and Hellas basin oceans. Our current model is not able to simulate details regarding the glacier flow, but only a global mass flux from land to the ocean. The simplified mass flux is estimated by restricting snow accumulation to 2 m of H$_2$O. Any excess is immediately redistributed over the ocean to maintain conservation of water in the system. Glacial processes such as accumulation (snow compaction), flow (path, erosion), and ablation (direct melting, fluvio-glacial river) are thus not included. Nevertheless, the global mass flux (see Materials and Methods) is estimated around $3\times10^{15}$ kg.y$^{-1}$ and seems to be very robust to obliquity changes and H$_2$ content. For comparison, this flux is currently 1 to $5\times10^{14}$ kg.y$^{-1}$ on the Earth today \cite{Radic_GlaciersEarthsHydrological_SiG2013}. In the Last Glacial Maximum, closer to our Martian cold and wet climate, the water discharge was estimated to $10^{16}$ kg.y$^{-1}$ for North America only \cite{Wickert_ReconstructionNorthAmerican_ESD2016}. Given the relative sizes of the planetary bodies, our value seems reasonable.

In the lowest altitude domain, called the "wet lowlands", rain, evaporation and surface runoff are balanced. Wet lowlands represent a minor surface fraction (22\% at 40$^\circ$ obliquity) of the planet for H$_2$=10\% (see Fig. \ref{Fig:Scheme-climate}). 

The rain occurs mainly over the ocean ($\sim$ 70\% of the total rain), and $\sim$ 60\% of the ocean evaporation is compensated by rain. The snow precipitation over the icy highlands is intense but $\sim$ 80\% is compensated by high sublimation rates. Our climate model differs from the "icy highlands model" \cite{Wordsworth_Globalmodellingearly_I2013, Palumbo_EarlyMarsClimate_GRL2018} since the ocean is stable due to glacier return flow.

From these results, one can describe a cold and wet climatic regime (see Fig. \ref{Fig:Scheme-climate} bottom) with significant glacier return flow from the cold highlands back to the ocean with moderate surface runoff near the shoreline.

\section*{Comparison to geological evidence}

\begin{figure}
\centering
\includegraphics[width=1.\linewidth]{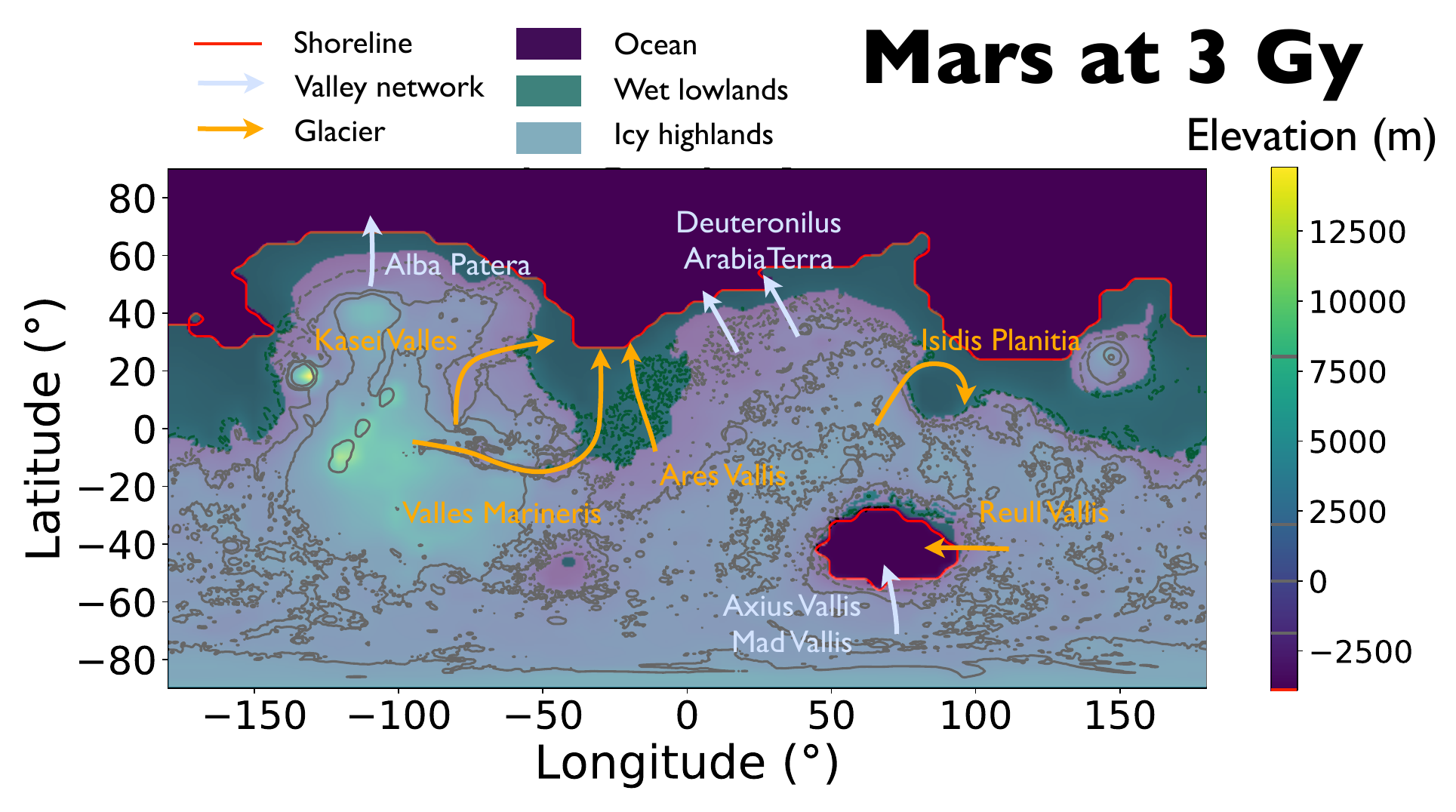}
\includegraphics[width=.8\linewidth]{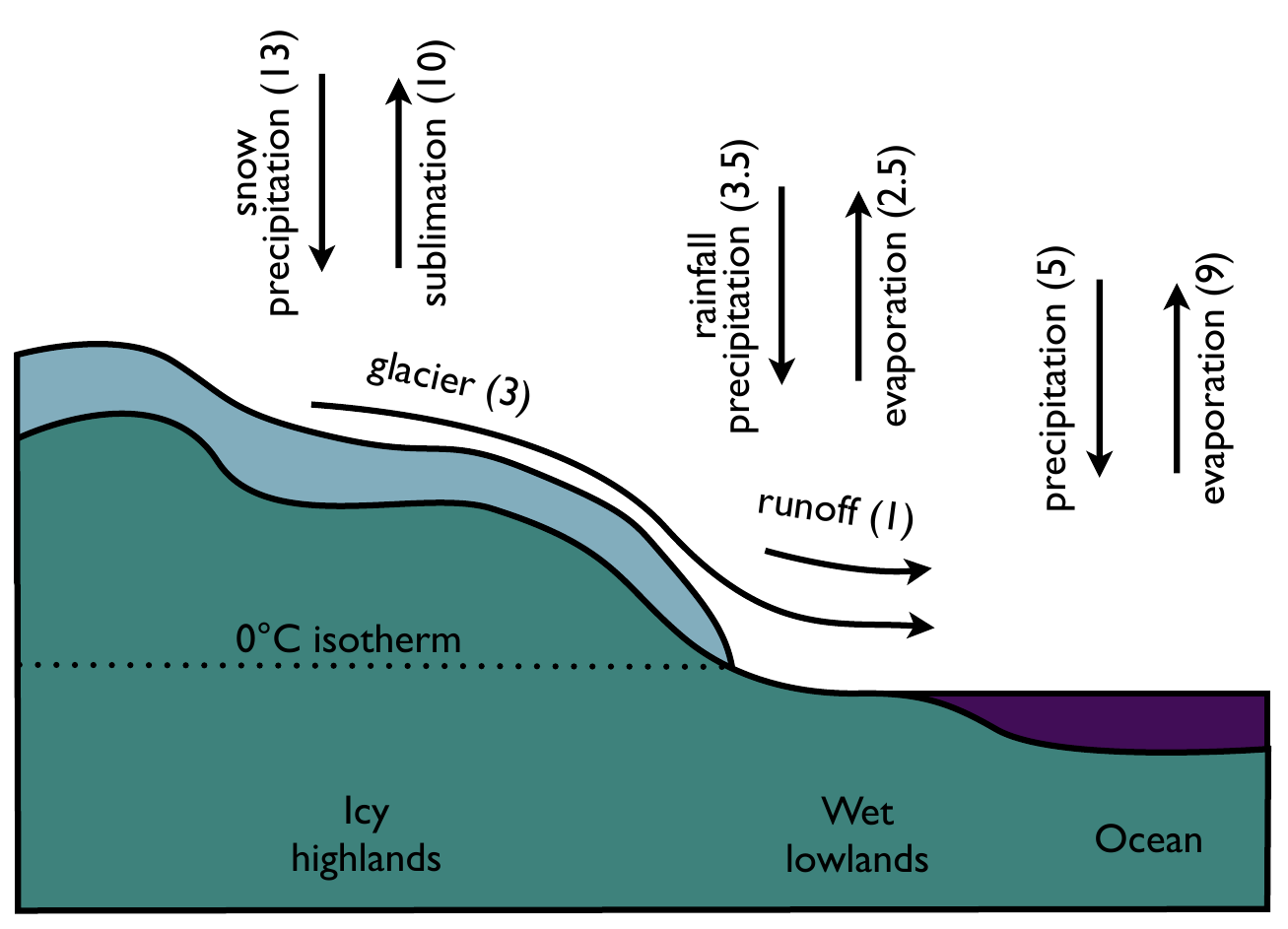}
\caption{Proposed scenario of a cold and wet climate at the Hesperian age (3 Ga). (top) Map of the climatic zones at 40$^\circ$ obliquity: ocean, wet lowlands and icy highlands, separated respectively by -3900 m and the 0$^\circ$C isotherm (-1990 m). The topography is modern topography without the North Polar Layered Deposits. Geological evidence of past glaciers with possible paths indicating return flow to the oceans are highlighted. Geological evidence of high drainage density valley networks 3 Ga is rare but noted above. (bottom) Simplified scheme of the hydrological cycle using box modeling. Fluxes in parenthesis have units of $10^{15}$ kg.y$^{-1}$.}
\label{Fig:Scheme-climate}
\end{figure}

The proposed scenario is in agreement with the long term stability of the ocean, with the paleo-shoreline and potential tsunami deposits. In addition, this climatic regime is in agreement with independent geological evidence concerning fluvial and glacial processes. Figure \ref{Fig:Scheme-climate} plots the major geological features of similar late Hesperian/early Amazonian age, not covered by subsequent units (Sup. Mat. section 2).

Contrary to the Noachian epoch, the modest appearance and development of dendritic valleys (rainfall related) during the late Hesperian/early Amazonian has been tied to major changes in climate. Several valleys at 2.9 Ga have been identified in the North of Alba Patera with a substantial temporal peak in the drainage density \cite{Hynek_Updatedglobalmap_JoGR2010}, located in the dichotomy boundary, in agreement with large rainfall predicted by our model (see Fig. \ref{Fig:GCM-output}). In addition a relatively high density network has been located in the South of Hellas basin \cite{Bernhardt_Photogeologicmappinggeologic_I2016}. Along the highland/lowland boundary (northwestern Arabia Terra and in Deuteronilus Mensae regions) network aligned channels are observed displaying streamlined islands of Hesperian age of possible marine \cite{Rodriguez_Tsunamiwavesextensively_SR2016} or fluvial origin \cite{Tanaka_GeologicmapMars_UGSR2014}. The wet lowlands have been substantially covered by more recent Amazonian processes (including volcanic resurfacing) \cite{Tanaka_GeologicmapMars_UGSR2014}, erasing most of the geological features of this age available at the surface.

Global extensive ice sheets have been proposed in the Southern hemisphere. Some deposits in the region of Malea Planum (Pityusa Patera) are possible remnants of an extensive polar ice sheet emplaced during the Hesperian period  \cite{Head_ExtensiveHesperianaged_JoGRP2001, Leonard_GeologicMapHellas_UGSR2001}. 
In addition, a 3 Ga old polythermal ice sheet should had covered the entire basin of Isidis Planitia with a maximum thickness of 4.9 km \cite{Soucek_3Gaold_EaPSL2015, Guidat_LandformassemblageIsidis_EaPSL2015}. Since this basin is close to the 0$^\circ$C isotherm, a glacier could have flowed from the accumulation zone in the highlands down to the ablation zone in Isidis, as mapped by J{\"o}ns \cite{Joens_LargeFossilMud_LPSC1987}. 

Reull Vallis is a 1500 km long U-shaped valley post-dating the Hesperian ridged plains with a long and complex history \cite{Mest_GeologyReullVallis_I2001}. It has been carved by 
glacial flow along the channel during a younger resurfacing episode of late Hesperian-early Amazonian age \cite{Kostama_Topographicmorphologiccharacteristics_JoGR2007}. 
With a length of $\sim$ 3000 km, $\sim$ 500 km width, and a depth of $\sim$ 3 km, Kasei Valles is the largest outflow channel on Mars. Flooding activity occurred mostly at the Hesperian between 3.6 and 3.8 Ga, with continuing activity possible to 2.5 $\pm$ 1 Ga. \cite{AndrewsHanna_Hydrologicalmodelingoutflow_JoGRP2007}. The morphology of Kasei Valles region is consistent with an origin by ice with a striking analogy to Alaskan glaciers \cite{Lucchitta_IcesculptureMartian_JoGR1982} and Antarctic ice streams \cite{Lucchitta_Antarcticicestreams_GRL2001}. A glacial interpretation of the Kasei Valles outflow was also proposed to explain the presence of scour features, tunnel valleys and esker forms \cite{Arfstrom_EquatorialIceSheets_2020}. Ares Vallis is a 1500 km long Hesperian outflow channel displaying multiple catastrophic flooding events. Its floor shows geomorphological evidence of ice-covered flooding \cite{Wallace_Evaporationiceplanetary_I1979} and glacial erosion \cite{Pacifici_GeologicalevolutionAres_I2009} with subsequent thermokarst degradation \cite{Costard_Thermokarstlandformsprocesses_G2001}. 
The presence of thermokarst lakes formed after flooding \cite{Costard_Thermokarstlandformsprocesses_G2001, Warner_Hesperianequatorialthermokarst_G2010} supports a hypothesis of the thawing of ground ice during the Hesperian (i.e., 3.6–3.0 Ga). In addition, evidence of past glaciations (between 1.4 Ga to 3.5 Ga) was reported within Valles Marineris. Morphological evidence shows extensive subglacial erosion of the lower parts of Valles Marineris wallslopes, together with the occurance of sackung deformations by deglaciation of the upper wallslopes \cite{Mege_EquatorialglaciationsMars_EaPSL2011, Gourronc_Onemillioncubic_G2014}.

Finally, one has to note that the extensive outflow channels are contemporaneous with the ocean.  An ocean at the Hesperian is also compatible with the timing of peak outflow channel activities \cite{Clifford_MarsHydrosphere_Icarus2001,Warner_refinedchronologycatastrophic_EaPSL2009}. Outflows probably formed by aquifer disruption \cite{Carr_FormationMartianflood_JoGR1979} or glacial erosion \cite{Lucchitta_IcesculptureMartian_JoGR1982,Pacifici_GeologicalevolutionAres_I2009}. More abundant signs of glaciation in the icy highlands may have been prevented by less erosive cold-based glaciers \cite{Fastook_EarlyMarsclimate_Icarus2012} and by post-dating Amazonian processes, such as volcanic events \cite{Tanaka_GeologicmapMars_UGSR2014}.

One also has to consider the arguments that confront the stable ocean scenario. First, the interpretation and location of the paleo shoreline itself is matter of controversy \cite{Carr_OceansMars_jgr_2003, Sholes_QuantitativeHighResolution_JoGRP2019}. Some studies claim that the putative shorelines are mutually inconsistent \cite{Sholes_WhereareMars_JoGRP2021} but it mainly concerns Contact 1 (Arabia potential shoreline), not Contact 2 (younger Deuteronilus shoreline) that we use in our study. Second, mineralogical data seems inconsistent with a Late Hesperian ocean \cite{Bibring_MarsHistory_Science2006,Carter_HydrousmineralsMars_JGRP2013}. In particular, mineralogy from impact crater excavated terrains \cite{Pan_stratigraphyhistoryMars_JoGRP2017} and clay occurrence in  central peaks \cite{Sun_Ancientrecentclay_JoGRP2015} are more consistent with a volcanic origin. But  widespread sedimentary rock has been identified in the Northern plains of Mars \cite{Salvatore_Evidencewidespreadaqueous_G2014}. Lastly, there is a lack of observed carbonates at the surface, that imply that the ocean was acidic. However, an acidic ocean seems unfavorable in presence of mafic rocks \cite{Niles_GeochemistryCarbonatesMars_SSR2012}.  Alternative scenarios to an ocean have been proposed implying lava flows \cite{Head_NorthernlowlandsMars_JoGR2002} or ice-related processes \cite{Lucchitta_MarsandEarth_I1981,Clifford_MarsHydrosphere_Icarus2001} creating the smooth northern plains we observe today. In addition, crater ejecta have been proposed as an alternative to tsunami deposits to form thumbprint terrains \cite{Kite_GeologicConstraintsEarly_SSR2019}.

\section*{Conclusion}

Our results demonstrate that a cold and wet climate could have been stable and is consistent with geomorphological evidence thanks to glacier return flow. Using fully coupled atmosphere / dynamic ocean modeling, we show that the ocean's circulation regionally warms the surface up to 4.5$^\circ$C. In these conditions the ocean is stable even for a global mean planetary temperature below 0$^\circ$C. This Martian climate may be similar to ancient Earth's with an active water cycle around the time of the early stages of life's appearance. Future work should encompass careful analysis of this stable ocean domain and its application to the past Martian climate, especially in the Noachian epoch. In addition, we hope that this work will stimulate more precise ice sheet simulations on Mars at global scale, in comparison with detailed photo-geomorphological studies. Further in situ analysis of the ocean boundary, for instance by Zhurong and ExoMars Rosalind Franklin rover, should confirm the shoreline and tsunami deposits interpretation. In the longer term, Mars Ice Mapper and particularly its high resolution radar capability will also be able to bring new clues.

\subsection*{Data Archival}
Model runs input and output will be available here:\url{https://portal.nccs.nasa.gov/GISS_modelE/ROCKE-3D/publication-supplements/}

\matmethods{

\subsection*{Atmosphere 3D-GCM}\label{sec:3DGCMDetails}

The 3-D General Circulation Model (3D-GCM) simulations performed herein utilize ROCKE-3D (R3D) \cite{Way2017}, which is based upon a parent Earth Climate Model called ModelE2 \cite{Schmidt2014}. ModelE2 has been used for the Coupled Model Intercomparison Project Phase 6 (CMIP6) \citep{Kelley2020}. R3D has a number of extensions in areas relative to non-Earth terrestrial worlds \cite{Way2017}. In particular R3D utilizes a separate and highly flexible radiative transfer scheme known as SOCRATES \cite{Edwards1996a, Edwards1996b} that allows for non-Earth atmospheric gas constituents and abundances which are outside the capabilities of the native ModelE2 scheme. R3D also extends the ModelE2 calendaring system and allows for worlds with different rotational and orbital parameters. In this project we use 4$^\circ\times5^\circ$ horizontal resolution for all components. The atmosphere has 40 layers with the model top at 0.1 hPa. The major components of surface and atmosphere exchange data at a time step of 1850 s, although shorter time steps are used internally. We set the solar luminosity to be $\sim$ 79\% \cite{Gough1981} of its current value (1360.67 W m$^{-2}$ at Earth) following \cite{Claire2012}, Hence at 3 Ga, the flux at Mars was set to 452.8 W m$^{-2}$. For this work we rely upon CO$_{2}$+H$_{2}$ CIA tables from \cite{Wordsworth2017} incorporated into R3D. Recent work \cite{Turbet_Measurementssemiempirical_I2020} has demonstrated the possibility that \cite{Wordsworth2017} may overestimate the warming provided in their CO$_{2}$+H$_{2}$ CIA calculations, confirmed by \cite{Godin2020}. At the same time the R3D SOCRATES pure CO$_{2}$ absorption is underestimated allowing $\sim$7.5 Wm$^{2}$ additional outgoing long wave radiation compared with SMART \cite{Kopparapu2013} calculations as shown in Figure 2 of \cite{Guzewich2021}.
The overestimated CO$_{2}$+H$_{2}$ CIA R3D uses from \cite{Wordsworth2017} tends to cancel out the underestimated CO$_{2}$ absorption as demonstrated in experiments conducted in \cite{Guzewich2021}. A more detailed explanation of these effects can be found in \cite{Guzewich2021}. However, even if we are underestimating the absorption we have ``bounded"  the problem in this work by providing GCM results for a 20\% H$_{2}$ atmosphere for the four different obliquities in the Supplementary Materials. 

H$_2$ provides a powerful greenhouse component in combination with CO$_2$ as a background gas, but other gas combinations involving CH$_4$ or H$_2$S may have an equivalent radiative effect at sufficient mixing ratios even if the motivation for their use is lacking. Recent measurements of collision induced absorption of CO$_2$ and CH$_4$ shows that this possibility is not favored \cite{Turbet_Measurementssemiempirical_I2020}. We run simulations with 10\% and 20\% H$_2$ in a CO$_2$ dominated atmosphere at 1 bar surface pressure to simulate the climate at 3 Ga, as in \cite{Turbet_paradoxesLateHesperian_SR2019}. We run simulations with 0$^\circ$, 20$^\circ$, 40$^\circ$, 60$^\circ$ obliquity, since this orbital parameter can have large excursions from the mean value in the past \cite{Laskar_INSOLATIONofMARS_Icarus_2004,Armstrong2004}.

H$_{2}$O cloud implementation in R3D is described in detail in \cite{Schmidt2014,Way2017,Guzewich2021}. The ice particles are assumed to be Mie scatterers while the parameterization of cirrus ice cloud properties follows \cite{Edwards2007} via our use of the SOCRATES radiation scheme in R3D (see Section \ref{sec:3DGCMDetails}). R3D does not include CO$_{2}$ ice clouds which are presently under development. For the relatively warm temperatures of our simulations they are expected to play a minor role in the radiation budget of the atmosphere, although for cooler climates they are known to play an important role \cite{Forget1997}.

The surface radiative properties of R3D are described in \cite{Schmidt_PresentDayAtmospheric_JoC2006}. Dry soil albedo is set to 15\%. The spectral and solar zenith angle dependence of ocean albedo is based on calculations of Fresnel reflection from wave surface distributions as a function of wind velocity \cite{Cox_Slopesseasurface_BotSIoO1956}. The effects of foam on ocean albedo are also included \cite{Gordon_Albedooceanatmospheresystem_AO1977}. In addition, R3D also includes the dependence of the cosine of the zenith angle for snow albedo both on ocean ice and on land \cite{Wiscombe_ModelSpectralAlbedo_JotAS1980}.

The main point of comparison with \cite{Kamada_coupledatmospherehydrosphereglobal_I2020} is the global average temperature for 20\% H$_{2}$. They found 15.65$^\circ$C (for us  16.2°C for slab ocean) for an obliquity of 25.19$^\circ$ (20$^\circ$ for us), and a solar flux 441.1 W.m$^{-2}$ (452.8 W.m$^{-2}$ for our later epoch). Given the relative numbers, we conclude that both models are in agreement.

\subsection*{Ocean Circulation Model}

The ocean shoreline is set to -3900 meters in all runs \cite{Head_PossibleAncientOceans_Science1999, Ivanov_TopographyDeuteroniluscontact_PaSS2017}. The bathymetry is assumed to be identical to present time except that the northern polar cap was removed. According to modern topography analysis, the Late Hesperian stage of the paleoshoreline is in agreement with current altitude, i.e.: without a correction from True Polar Wander due to Tharsis \cite{Head_PossibleAncientOceans_Science1999,Perron_Evidenceancientmartian_N2007,Citron_Timingofoceans_N2018,Bouley_LateTharsisformation_N2016}. The ocean horizontal resolution
is the same as the atmosphere: 4$^\circ$ $\times$ 5$^\circ$ and can extend to 13 layers or 4647m. The ocean bathymetry is determined by the topography. In Hellas basin the ocean depth reaches 3194 m (level 12 of 13) while in the northern ocean its deepest extent is 1294 m (level 10 of 13). The mesoscale diffusivity is fixed at 1200 m$^{2}$s$^{-1}$. At model start the ocean is initialized as liquid without any sea ice. 

For all simulations herein the dynamic ocean model assumes Earth's values of salinity (35 PSU) at model start. The ocean includes a number of additional dynamical processes that change as the model moves forward in time such as sea ice growth/shrinkage, salinity change due to fresh water incorporation by surface runoff, precipitation/evaporation and associated changed in sea height etc. Additional details on the ocean model can be found in \cite{Schmidt2014,Way2017,DelGenio2019}. 

The ocean surface fraction is $\sim$16\% which is small in comparison to Earth at $\sim$ 71\%. The ocean is also shallower than the mean depth for Earth. For this reason the time to bring our ancient Mars model ocean and atmosphere into equilibrium is much shorter ($\sim$ 100s of years) than would be the case for an Earth like ocean ($\sim$ 1000s of years). We assume this equilibrium has been reached when the net radiative balance (the difference between incoming and outgoing fluxes) is less than 0.2 W m$^{-2}$ (see Sup. Mat. section 1.5). Model runs with the dynamic ocean typically require 300-400 martian years to reach equilibrium, although in some cases a checkpoint-restart file from a similar experiment was used to start the simulation in a condition close a predicted end state. For the latter such simulations may only need 100 years to reach equilibrium. All outputs are averaged over 10 Martian years after radiative equilibrium is attained, and these are used for most of the analysis in this paper. All quantities are expressed in Earth years, noted "y". When referring about time before present, we used the Earth year ago, noted "a".

\subsection*{Land hydrological cycle}\label{sec:land-hydro}

Land surface hydrology \cite{Schmidt_PresentDayAtmospheric_JoC2006,Rosenzweig1997} is represented by 6 layers of soil (up to a depth of 3.5 m), a 3-layer snow model and dynamic lakes. The land receives water in the form of precipitation (either rain or snow) and loses it through evaporation/sublimation and via runoff. For the land surface martian regolith is approximated by sand (with equivalent porosity, hydraulic conductivity and thermal inertia), initially saturated with 90\% liquid water (representing 1 m GEL). The surface runoff depends on the strength of the rain and the level of upper soil layer saturation, while the underground runoff from deeper layers depends on the amount of water in the layer and local slope (in the present simulations the slope was set uniformly to a typical flat desert value). The runoff water goes directly to the lakes, which expand or shrink depending on the amount of available water and exchange water with the lakes in neighboring cells according to a prescribed river routing scheme based on topography. In addition to sublimation, the snow model incorporates the algorithms of melting and meltwater refreezing in deeper layers as well as gravitational compaction of the layers. There is no transport of snow between the cells, i.e. we don’t model the glacier dynamics. It is challenging to model glaciers inside a GCM due to their much longer time-scales ($\sim$10$^{4-5}$ years) versus the typical timescales of GCM simulations ($\sim$10$^{3}$). In CMIP studies ice sheets are normally calculated off-line and incorporated into GCM simulations at different epochs \cite{Abe-Ouchi2015}. Here we use a simplified approach, also used to model Earth climate, for instance the Last Glacial Maximum, 21,000 years ago \cite{Kageyama_GMD2017}. We assume that the areas of permanent presence of snow in our simulations represent glacier heads. We assume that in steady state any additional accumulation of snow over these regions is compensated by ice flow to the lower altitudes where it will either melt or eventually reach the ocean. At the typical surface temperatures in our simulations ($>$ 250K), the ice can easily flow on Mars \cite{Colaprete_JGR1998, Lucchitta_GRL2001, Fastook_EarlyMarsclimate_Icarus2012}. Since we are not modeling the thickness of glaciers, there is no reason to allow snow to accumulate to very high depth. This will require much longer spin up times but have no noticeable effect on results. So we restrict the snow thickness to 2 m of H$_2$O equivalent ($\sim$ 3.5m of compacted snow), which is enough to properly simulate its effect on climate. Any accumulated snow in excess of this amount is immediately removed from the snowpack and is redistributed over the ocean to maintain conservation of water in the system.

\subsection*{Box modeling}

The complex 3D climate system is summarized by box modeling. We divided the surface into 3 zones : icy highlands, wet lowlands and ocean, delimited respectively by the mean altitude of the 0$^\circ$C isotherm and -3900 m. For each surface box, but also for the atmosphere and the ocean, we compute the sum of all incoming and outgoing fluxes of water, including snow/rain precipitation, evaporation, surface runoff and simplified glacier return flow (See Sup. Mat. sections 1.3 and 1.4). We use sand (see Section \ref{sec:land-hydro}) for our soil and it is initially saturated with 90\% liquid water. This saturation amount will change according to climatic conditions during the course of the simulation (e.g. precipitation or evaporation). The timescale for this process to reach equilibrium can be quite long ($\sim$ 1000s of years). Since we are not interested in this process, we estimate and remove this source of water flux in our box model. We check that the hydrological cycle is in balance by examining related diagnostics in the model, but there are unaccounted for fluxes at $<= 5\%$ due to incomplete GCM diagnostics, from salinity changes and numerical rounding. The supplementary material presents the mass fluxes for all model runs.
}

\showmatmethods{} 

\acknow{We acknowledge support from the “Institut National des Sciences de l'Univers” (INSU), the "Centre National de la Recherche Scientifique" (CNRS) and "Centre National d'Etudes Spatiales" (CNES) 
through the "Programme National de Plan{\'e}tologie". M.J.W. was supported by the National Aeronautics and Space Administration (NASA) Astrobiology Program through collaborations arising from his participation in the Nexus for 
Exoplanet System Science (NExSS) and the NASA Habitable Worlds Program. Resources supporting this work were provided by the NASA High-End Computing (HEC) Program through the NASA Center for Climate Simulation (NCCS) at Goddard Space Flight 
Center. M. J. W. and A. I. acknowledge support from the GSFC Sellers Exoplanet Environments Collaboration (SEEC), which is funded by the NASA Planetary Science Division's Internal Scientist Funding Model.}

\showacknow{} 

\bibliography{reference}

\end{document}